%% file: head.tex
\documentclass[sigconf]{acmart}
\acmDOI{x}

\usepackage{booktabs} 
\usepackage{graphicx}
\usepackage{ifthen}

\usepackage[utf8]{inputenc} 
\usepackage[T1]{fontenc}    
\usepackage{url}            
\usepackage{booktabs}       
\usepackage{amsfonts}       
\usepackage{dsfont}

\usepackage{nicefrac}       
\usepackage{microtype}      
\usepackage{amsmath}
\usepackage{bm}
\usepackage{subcaption}
\usepackage{mathrsfs}
\usepackage{xcolor}
\usepackage{enumitem}
\usepackage{algorithm}
\usepackage{algorithmic}
\usepackage{multirow}
\usepackage{hyperref}
\hypersetup{colorlinks,allcolors=black}

\usepackage{balance}

\newcommand {\out}[1]{}

\newcommand{\changed}[1]

\begin{document}
\title{Alleviating Cold-Start Problems in Recommendation\\through Pseudo-Labelling over Knowledge Graph}

\author{Riku Togashi}
\affiliation{
  \institution{CyberAgent, Inc.}
  \city{Tokyo} 
  \country{Japan}
}
\email{rtogashi@acm.org}

\author{Mayu Otani}
\affiliation{
  \institution{CyberAgent, Inc.}
  \city{Tokyo} 
  \country{Japan}
}
\email{otani_mayu@cyberagent.co.jp}

\author{Shin'ichi Satoh}
\affiliation{
  \institution{CyberAgent, Inc.}
  \city{Tokyo} 
  \country{Japan}
}
\email{satoh@nii.ac.jp}

\renewcommand{\shortauthors}{
}

\begin{abstract}
Solving cold-start problems is indispensable to provide meaningful recommendation results for new users and items.
Under sparsely observed data, unobserved user-item pairs are also a vital source for distilling latent users' information needs.
Most present works leverage unobserved samples for extracting negative signals.
However, such an optimisation strategy can lead to biased results toward already popular items by frequently handling new items as negative instances.
In this study, we tackle the cold-start problems for new users/items by appropriately leveraging unobserved samples.
We propose a knowledge graph (KG)-aware recommender based on graph neural networks, which augments labelled samples through pseudo-labelling.
Our approach aggressively employs unobserved samples as positive instances and brings new items into the spotlight.
To avoid exhaustive label assignments to all possible pairs of users and items,
we exploit a KG for selecting probably positive items for each user.
We also utilise an improved negative sampling strategy and thereby suppress the exacerbation of popularity biases.
Through experiments, we demonstrate that our approach achieves improvements over the state-of-the-art KG-aware recommenders in a variety of scenarios; in particular, our methodology successfully improves recommendation performance for cold-start users/items.
\end{abstract}


\maketitle

\input{body}
\bibliographystyle{ACM-Reference-Format}
\bibliography{head.bib}

\end{document}

%% file: body.tex
\section{Introduction\label{sec_intro}}
Recommendation systems aim to capture users' interests based on various kinds of clues and help users discover new items.
The primary source in personalised recommendation tasks is implicit user feedback (e.g. clicking, watching, and purchasing), which is routinely logged and reflects users' interests.
However, even with such abundant users' feedback,
recommender systems still suffer from the \emph{cold-start problem};
for new users who have not yet interacted with enough items,
recommender systems inevitably experience a lack of information.
The problem is severe in applications where the user and item databases are frequently updated, such as e-commerce, social news media, and online advertisement.
New items are also a typical cause of the cold-start problem, particularly in social networking services where user-generated contents continuously come in.
In such applications, recommender systems are required to handle new items associated with few interactions.

One option to mitigate this problem is to introduce knowledge graphs (KGs), which provide side-information about items~\cite{zhang2016collaborative,wang2018dkn,huang2018improving,yu2014personalized,zhao2017meta,MCRec}.
KGs are heterogeneous graphs of entities and relations between entities.
The nodes represent entities (e.g. items and their properties),
and the edges represent the relations between entities.

The problems due to lack of information remain challenging even with KGs.
Interactions observed in the form of implicit feedback provide only positive signals, and
this causes the fundamental obstacle in recommendation tasks, called the one-class problem~\cite{pan2008one,hu2008collaborative}.
For handling this only-positive setting, unobserved samples are the key resources, particularly for the sake of optimisation.
Conventional methods leverage unobserved samples to extract negative signals,
relying on the assumption that the unobserved samples can be utilised as negative instances~\cite{pan2008one,BPRMF,he2016fast}.
However, as this strategy distils only negative signals from unobserved samples,
it may lead to biased and sub-optimal results;
recommender systems may underestimate the chance of positive feedback of users or items rarely observed, namely, cold-start users and items~\cite{fairnessZuohui20}.

KG-aware recommender systems that utilise graph neural networks (GNNs)~\cite{wang2019knowledge,KGAT}
are promising directions.
GNNs leverage a KG in an end-to-end manner and
involve unobserved samples in the training phase by propagating features from labelled nodes to unlabelled nodes over a KG.
This nature of GNNs is suitable for the transductive setting of recommendation problems wherein
the goal is to predict labels for unlabelled nodes that are known in the training phase.
Nevertheless, conventional methods leverage unobserved samples mainly as negative instances based on the aforementioned assumption.

In this paper,
we address to improve recommendation for new items and users without compromising overall performance by appropriately leveraging unobserved samples.
We propose a KG-aware recommender based on GNNs and pseudo-labelling, namely, \emph{KGPL}.
Instead of handling unobserved samples only as negative,
we assign pseudo-labels to unobserved samples by model predictions that reflect the knowledge learnt from the observed data and KG.
Hence, in KGPL, unobserved samples can be handled as weak-positive instances in contrast to the conventional methods.
To ensure the reliability of pseudo-labels,
we carefully select the unobserved items to be labelled
through two sampling strategies;
(1) KG-aware sampling of items for pseudo-labelling based on the graph structures rooted at users and observed items in a KG; and
(2) popularity-aware sampling of items for negative instances.
Our KG-aware sampling for pseudo-labelling selects probably positive items
for a given user while ensuring the reliability of pseudo-labels.
We also introduce a negative sampling strategy for enhancing the stability of training.
Besides, to achieve further robust optimisation,
we propose a co-training approach that utilises a pair of GNNs.
Experimental results demonstrate that our method is effective for a wide range of users with different levels of data sparsity.

As overall average measures cannot describe the detailed difference between methods,
we showcase a series of experiments to extensively compare the behaviour of KG-aware recommenders with assuming user-side and item-side cold-start situations.
Our experiments provide more detailed insights on recommenders' behaviours which conventional evaluation methodology roughly abstracts.

\section{Related Work}
\subsection{Recommendations with Knowledge Graphs}
For leveraging a KG as additional side-information about items, various approaches are examined, such as
embedding-based methods~\cite{zhang2016collaborative,wang2018dkn,huang2018improving,wang2019multi},
path-based methods~\cite{yu2014personalized,zhao2017meta,hu2018leveraging,lu2020meta}, and
hybrid methods~\cite{RippleNet,sun2018recurrent,wang2019knowledge2,KGAT}.
Embedding-based methods alleviate cold-start problems 
by extracting semantic knowledge from a KG.
Path-based methods manually design patterns of connectivity among items (i.e. meta-paths or meta-graph) to extract user-specific relatedness.
Recent hybrid methods adopt graph neural networks (GNNs) to learn representations of entities and user-specific relatedness without labour-intensive meta-paths~\cite{wang2019knowledge,wang2019knowledge2,KGAT}.
However, as in traditional recommender systems such as collaborative filtering~\cite{hu2008collaborative}, these methods leverage unobserved samples only as negative instances during training.
Wang \emph{et al.} proposed a negative sampling strategy based on reinforcement learning~\cite{KGPolicy20}.
Their method mainly focuses on finding informative negative samples and does not concern the problem of cold-start items that are rarely observed but can be relevant to certain users.
Our method utilises unobserved samples as both positive and negative instances to alleviate the cold-start problem.
To pull up cold-start items,
our KG-aware sampling and pseudo-labelling approach extract probably positive instances from unobserved samples.

\subsection{Graph-Based Semi-Supervised Learning and Pseudo-Labelling}
Semi-supervised learning (SSL) aims to correctly label all samples when a small fraction of samples are labelled.
A recommendation task with implicit feedback can be considered as an SSL problem;
in particular, the setting can be regarded as a transductive learning task wherein the unlabelled samples (i.e. user-item pairs) for inference are known in the training phase.
GNNs handle such a transductive setting
by propagating features over a graph~\cite{bruna2013spectral,defferrard2016convolutional,kipf2016semi,duvenaud2015convolutional,niepert2016learning,hamilton2017inductive}.
The recent success of GNN-based recommenders also demonstrates the effectiveness of GNNs~\cite{ying2018graph,monti2017geometric,berg2017graph,wang2019knowledge2,KGAT}.

Graph-based SSL has been extensively explored in various fields.
Existing methods often rely on
assumptions on the distribution of labels over a graph, such as label smoothness~\cite{wang2007label,karasuyama2013manifold,wang2019knowledge2}.
By adopting such assumptions, label propagation is widely utilised in a transductive setup~\cite{wu2012learning,Douze_2018_CVPR,iscen2019label}.
However, applying label propagation techniques to KG-aware personalised recommendation is not trivial for three reasons;
(1) only positive instances can be observed in implicit feedback settings;
(2) labels and edge weights depend on users' taste, and therefore, edges may not indicate similarity between their connecting nodes; and 
(3) labelled nodes (i.e. the source of label propagation) for a single user are scarce.
Wang \emph{et al.} proposed a KG-aware recommender based on GNNs
with a regularisation technique that induces label smoothness in model predictions, namely, KGNN-LS~\cite{wang2019knowledge2}.
Their proposed regularisation technique realises label propagation on a KG through the \emph{leave-one-out} loss optimisation that
holds out an observed item for a user and leverages it as an unlabelled instance.
However, as they utilise only observed samples for their regularisation technique,
and unobserved samples are still not leveraged explicitly as positive.

Pseudo-labelling has been adopted particularly by deep learning-based methods~\cite{lee2013pseudo,shi2018transductive,ke2020multi}.
It reuses repeatedly updated model predictions as true labels for training, and thereby directly leverages unobserved samples as training data.
However, pseudo-labels generated by a model may be noisy, and leveraging all samples is infeasible in the context of recommendation problems as the number of all possible pairs of users and items can easily reach billion-scale.

Our method employs pseudo-labelling but inherits the spirit of label propagation.
To avoid exhaustively labelling unobserved samples,
we exploit graph structures for selecting candidates that can be labelled reliably.
In contrast to KGNN-LS, our loss function involves unobserved samples also as positive instances.

\section{Proposed Method\label{method}}
\subsection{Problem Formulation}
We now explain the primary setting of the problem discussed in this paper.
In a scenario of personalised recommendation,
we have a set of users $\mathcal{U}$ and a set of items $\mathcal{I}$.
The user-item interaction matrix $\bf{Y}$ can be obtained as the users' implicit feedback; we let $y_{u,i}$ be the $(u,i)$-entry of $\bf{Y}$, and then $y_{u,i}=1$ indicates that user $u$ has engaged with item $i$.
Unfortunately, we cannot observe $\bf{Y}_{ui}$ for all user-item pairs $(u, i) \in \mathcal{U}\times\mathcal{I}$ in realistic situations;
thus, $\bf{Y}_{ui}$ can be missing for the majority of $(u, i)$.
Hence, we let $\mathcal{I}_{u}^{+}$ denote the set of observed items for $u$,
and $\mathcal{I}\setminus\mathcal{I}_{u}^{+}$ indicates the set of unobserved items for $u$.
We also let $\mathcal{O}=\{(u,i)|u \in \mathcal{U}, i \in \mathcal{I}_u^{+}\}$ denote the set of observed user-item pairs.
Moreover, in the knowledge-aware recommendation problem,
we have a KG $\mathcal{G}=\{(h, r, t)|h, t\in \mathcal{E}, r\in\mathcal{R}\}$,
where each triplet $(h, r, t)$ describes that there is a relationship $r$ between head entity $h$ and tail entity $t$;
$\mathcal{E}$ and $\mathcal{R}$ denote the sets of entities and relations, respectively.
Here, $\mathcal{E}$ comprises items $\mathcal{I} (\mathcal{I} \subseteq \mathcal{E})$ and non-item entities $\mathcal{E}\setminus\mathcal{I}$ (e.g. nodes corresponding to the attributes of items).

The focus of this paper is
predicting the $y_{u,i}$ for each unobserved sample $(u, i) \in (\mathcal{U}\times\mathcal{I})\setminus\mathcal{O}$
based on the observed interactions $y_{u,i}$ corresponding to $\mathcal{O}$.
To this end, we aim to obtain a prediction function $\hat{y}_{u,i}=\mathcal{F}(u,i|\Theta,\bf{Y},\mathcal{G})$,
where $\hat{y}_{u,i}$ denotes the probability that user $u$ will engage with item $i$, and $\Theta$ are model parameters of function $\mathcal{F}$. 

\subsection{Knowledge-Aware Graph Neural Networks\label{GNN-method}}
This section describes our GNN architecture for capturing users' preferences by exploiting a KG.
The architecture of our GNN is almost the same as KGNN-LS~\cite{wang2019knowledge2}.

We first model the user-specific relation scoring function that measures the importance of each relation for a user.
The $(j, k)$-th entry of a user-specific adjacency matrix ${\bf A}_u \in \mathbb R^{|\mathcal E| \times |\mathcal E|}$ can be obtained as follows:
\begin{eqnarray}
  \label{adj}
  (\bf{A}_u)_{jk} = \exp({\bf u}^{\top}r_{e_j, e_k})
\end{eqnarray}
where $\bf{r}_{e_j, e_k}$ is the relation between entities $e_j$ and $e_k$ in $\mathcal G$.
We set $(A_u)_{jk}=0$ if there is no relation between $e_j$ and $e_k$.
Then, the layer-wise forward propagation can be written by:  
\begin{equation}
  \label{eq:kgcn}
  {\bf H}_{l+1} = \sigma_{l}\left({\bf D}_u^{-1/2} \left( {\bf A}_u + {\bf I} \right) {\bf D}_u^{-1/2} {\bf H}_l {\bf W}_l \right), \ l = 0, 1, \ldots, L-1,
\end{equation}
where ${\bf H}_l$ is the matrix of the hidden representations of entities in the $l$-th layer, and ${\bf H}_0$ is a trainable parameter ${\bf E}$.
Here, ${\bf A}_u$ aggregates the representation vectors of neighbouring entities, and we introduce a self-connection loop by ${\bf A}_u + {\bf I}$.
We also let ${\bf D}_u$ denote a diagonal degree matrix with entries $(\bf{D}_u)_{jj} = \sum_k ({\bf A_{u}})_{jk}$ to normalise ${\bf A}_u$.
Here, ${\bf W}_l \in \mathbb R^{d_l \times d_{l+1}}$ is the layer-wise trainable weight matrix, and $L$ is the number of layers.
$\sigma_l$ is the activation function for the $l$-th layer, and
we adopt LeakyReLU~\cite{maas2013rectifier} for the intermediate layers ($l < L$) and $\tanh(\cdot)$ for the last output layer.

The predicted engagement probability of user $u$ with item $i$
is computed by $\hat{y}_{u,i} = \sigma({\bf u}^{\top} {\bf i}_u)$,
where ${\bf i}_u$ (i.e. the row corresponding to item $i$ in ${\bf H}_L$) is the final personalised representation vector of $i$,
and $\sigma(\cdot)$ is a sigmoid function.
In the training phase, we aim to optimise model parameters $\Theta=\{{\bf U}, {\bf E}, {\bf R}, {\bf W}_{0},\dots,{\bf W}_{L}\}$.
We apply dropout to ${\bf U}$ and ${\bf H}_l\,(l=1,\dots,L)$ with a same ratio instead of L2 regularisation.

\subsection{Semi-Supervised Learning Based on Pseudo-Labelling\label{pl-section}}
Our method leverages a high coverage of unobserved user-item pairs to optimise the model
without assuming all unobserved pairs as negative instances.
Our approach comprises mainly three techniques;
(1) pseudo-labelling framework for one-class settings (described in this section);
(2) sampling strategies for pseudo-labelled and negative instances (described in Section~\ref{sampling-opt}); and
(3) co-training for robust optimisation with noisy pseudo-labels as in Section~\ref{co-training}.

To alleviate the sparsity issues of observed interactions,
we augment positive and negative labelled data by predicting the labels
of unobserved samples.
Our label function for user $u$ and item $i$ can be written as follows:
\begin{eqnarray}
\label{label-function}
  \hat{l}_u(i) = \begin{cases}
    1, & \text{if $u \in \mathcal{U} \land i \in \mathcal{I}_u^{+}$},\\
    \hat{y}_{u,i}, & \text{if $u \in \mathcal{U} \land i \in \mathcal{I}_u^{\pm}$},\\
    0, & \text{if $u \in \mathcal{U} \land i \in \mathcal{I}_u^{-}$},\\
    \end{cases}
\end{eqnarray}
where $\mathcal{I}_u^{+}$, $\mathcal{I}_u^{-}$, and $\mathcal{I}_u^{\pm}$ denote the sets of positive, negative, and pseudo-labelled items for $u$, respectively.
Then, our loss function can be expressed as follows:
\begin{eqnarray}
\notag
  \mathcal{L} 
  &=& \sum_{u \in \mathcal{U}}\sum_{i \in \mathcal{I}_u^{+}}\ell\left(1, \hat{y}_{u,i};\Theta\right)
  +\sum_{u \in \mathcal{U}}\sum_{i \in \mathcal{I}_u^{\pm}}\alpha_{u,i}\ell\left(\hat{l}_u(i), \hat{y}_{u,i};\Theta\right)\\
  &+& \sum_{u \in \mathcal{U}}\sum_{i \in \mathcal{I}_u^{-}}\beta_{u,i}\ell\left(0, \hat{y}_{u,i};\Theta\right),
  \label{loss-function}
\end{eqnarray}
where $\ell(y_{u,i}, \hat{y}_{u,i};\Theta)$ is the loss function such as squared error and cross-entropy loss under a model with parameters $\Theta$.
$y_{u,i}$ and $\hat{y}_{u,i}$ denote a target label and a predicted label, respectively.
$\beta_{u,i}$ and $\alpha_{u,i}$ are the weight for the $(u, i)$-entry as a negative or pseudo-labelled instance, respectively.
We adopt the cross-entropy loss function as $\ell(\cdot)$.
As model prediction $\hat{y}_{u,i}$ can be considered as the probability that $i$ is positive for $u$,
the loss for a pseudo-labelled sample can be considered as an expected loss for the unobserved label;
$\ell(\hat{l}_{u}(i), \hat{y}_{u,i};\Theta)=-\hat{l}_u(i)\log(\hat{y}_{u,i})-(1-\hat{l}_u(i))\log(1-\hat{y}_{u,i})=\mathbb{E}_y[\ell(y, \hat{y}_{u,i};\Theta)]$.

There are two main challenges in the minimisation of $\mathcal{L}$;
(1) computing $\mathcal{L}$ is generally infeasible as $|\mathcal{U}\times\mathcal{I}|$ can easily reach the scale of billions in real applications; and
(2) training a model relying on probably noisy pseudo-labels may lead to poor optimisation.
To solve these obstacles, we propose a sampling-based learning framework for optimising $\mathcal{L}$.
As the unobserved samples dominate the whole of the user-item pairs,
we aim to efficiently select $\mathcal{I}_u^{-}$ and $\mathcal{I}_u^{\pm}$
for each user $u$ through two sampling strategies.

The expectation of loss function on a mini-batch, $\mathcal{L}_B$, can be expressed as follows:
\begin{eqnarray}
  \notag
  \mathbb{E}\left[\mathcal{L}_B\right]
    = \mathbb{E}_{u \sim Uni(\cdot)}\Bigg[&\,&\mathbb{E}_{i_{+}\sim P^{+}(\cdot|u)}\big[\ell(1,\hat{y}_{u,i_{+}};\Theta)\big] \\\notag
                                     &+&\mathbb{E}_{i_{-}\sim P^{-}(\cdot|u)}\big[\ell(0,\hat{y}_{u,i_{-}};\Theta)\big] \\
                                     &+&\mathbb{E}_{i_{\pm}\sim P^{\pm}(\cdot|u)}\big[\ell(\hat{l}_u(i_{\pm}),\hat{y}_{u,i_{\pm}};\Theta)\big]\Bigg]
\end{eqnarray}
where $P^{+}(i_{+}|u)$, $P^{-}(i_{-}|u)$ and $P^{\pm}(i_{\pm}|u)$ are the conditional distributions of the observed, unobserved, and pseudo-labelled items for $u$, respectively.
We set the distribution of users in the loss function to a uniform distribution so that each user contributes equally.
For observed data, we also utilise a uniform distribution as $P^{+}(i_{+}|u)$ to sample an observed item for a user.
The density of $P^{-}(i_{-}|u)$ and $P^{\pm}(i_{\pm}|u)$ are corresponding to weights $\beta_{u,i_{-}}$ and $\alpha_{u,i_{\pm}}$, respectively.

\subsection{Sampling Based on Knowledge Graph and Popularity\label{sampling-opt}}
\subsubsection{KG-Aware Item Sampling for Pseudo-Labelling\label{kgpl}}
To augment accurately labelled items for cold-start users through pseudo-labelling,
we sample items based on the graph structure centred on the observed user-item pairs.
The key idea to sample the items for pseudo-labelling is that,
for the items that can be reached through meta-paths (i.e. the paths rooted at a user's interacted items in a KG),
we can assign reliable pseudo-labels to the items for the user based on the knowledge learnt from observed samples; moreover, such items are probably positive items as they share entities with the positive items for the user.

To find items which we can assign reliable personalised pseudo-labels,
we count the number of paths to an item from the positive items for a user. 
Suppose that $\mathcal{I}_u^{+}$ is the set of observed items that $u$ has interacted with.
We list a set of item-to-item paths from an observed item $i_{+} \in \mathcal{I}_u^{+}$ to an unobserved item $i_{\pm} \in \mathcal{I}\setminus\mathcal{I}_u^{+}$ as $T_{i_{+}i_{\pm}}$.
We employ the $h$-hop breadth-first search (BFS) for searching the paths rooted at the observed items for a user. 
Based on these paths, we sample an unobserved item $i_{\pm}$ for $u$ according to the sampling distribution $P^{\pm}(i_{\pm}|u)$ that has the following probability density:
\begin{eqnarray}
  q(i_{\pm}|u) = \frac{n_{u,i_{\pm}}^{a}}{\sum_{i \in \mathcal{I}\setminus\mathcal{I}_u^{+}} n_{u,i}^{a}}, \,\,\,\, n_{u,i_{\pm}} = \sum_{i_{+} \in \mathcal{I}_u^{+}}|T_{i_{+}i_{\pm}}| 
\end{eqnarray}
where $n_{u,i_{\pm}}$ is the number of paths,
and $a \in \mathbb{R}$ is a hyper-parameter to control the skewness of the sampling distribution; $P^{\pm}(i_{\pm}|u)$ becomes close to a uniform distribution with small $a$.
To ensure the coverage of item candidates to be labelled, we set $n_{u,i}=0.5$ for item $i$ that is unreachable through the $h$-hop BFS.

\subsubsection{Popularity-Aware Negative Sampling\label{popns}}
Uniformly sampling negative items as in conventional works~\cite{RippleNet,wang2019multi,wang2019knowledge2,KGAT} can reinforce biases toward the popularity in practical settings.
Because negative sampling extracts unobserved items as a negative instance for a user,
cold-start items, i.e., items with few interactions, are unobserved items for most users.
Thus, cold-start items are more likely to be selected by negative sampling than popular items. 
This bias toward popularity can hurt performance, particularly for cold-start items and, thus, the quality of pseudo-labels for cold-start items.
In light of this consideration, we adopt a frequency-based strategy for negative sampling distribution $P^{-}(i_{-}|u)$,
as in conventional methods~\cite{rendle2014improving,he2016fast}.
We sample negative items from $\mathcal{I}\setminus\mathcal{I}_u^{+}$ according to the following probability:
\begin{eqnarray}
  \label{popns-density}
  p(i_{-}|u) = \frac{m_{u,i_{-}}^{b}}{\sum_{i \in \mathcal{I}\setminus\mathcal{I}_u^{+}}m_{u,i}^{b}}
\end{eqnarray}
where $m_{u,i_{-}}$ is the number of observed interactions for item $i_{-}$,
and $b$ controls the importance of frequency.

\subsubsection{Mini-Batch Sampling Strategy\label{mb-sampling}}
In the training phase, we construct mini-batches by utilising the aforementioned sampling distributions.
It should be noted that we sample an unobserved pair and determine its pseudo-label on the fly; 
each set of unobserved samples for pseudo-labelling is generated after sampling a mini-batch from the observed data $\mathcal{I}_u^{+}$.
Moreover, we ensure that a mini-batch contains three samples for a user, namely, positive, negative, and pseudo-labelled samples
to ensure the class balance in a mini-batch for a user.
Algorithm~\ref{alg:kapl} describes our mini-batch sampling strategy.

\begin{algorithm}[t]
  \caption{Mini-batch sampling algorithm}
  \label{alg:kapl}
  \begin{algorithmic}[1]
    \REQUIRE{Interaction matrix $\bf Y$, KG $\mathcal G$, Model $f$}
    \ENSURE{Mini-batch $\mathcal B$}
    \STATE $\mathcal B \leftarrow \emptyset$      
    \FOR{1/2 number of samples in a mini-batch}
      \STATE Sample a user by $u \sim Uni(\cdot)$
      \STATE Sample an observed item by $i_{+} \sim P^{+}(\cdot|u)$
      \STATE Sample a negative item by $i_{-} \sim P^{-}(\cdot|u)$;
      \STATE Sample an item for pseudo-labelling by $i_{\pm} \sim P^{\pm}(\cdot|u)$
      \STATE Compute pseudo-label $\hat y_{u,i_{\pm}}$
      \STATE Assign the pseudo-label as $\hat{l}_{u,i_{\pm}} \leftarrow \hat y_{u,i_{\pm}}$
      \STATE $\mathcal B \leftarrow \mathcal B \cup \{(u, i_{+}, 1), (u, i_{-}, 0), (u, i_{\pm}, \hat{l}_{u,i_{\pm}})\}$
      \ENDFOR
    \RETURN $\mathcal B$
  \end{algorithmic}
\end{algorithm}

\subsection{Co-training for Noisy Pseudo-Labels\label{co-training}}
Training a model on pseudo-labels that were generated by the model itself can make the optimisation unstable.
To realise more robust optimisation, we introduce a co-training approach~\cite{blum1998combining,qiao2018deep,han2018co}.
In our co-training approach, we train two models, $f$ and $g$, while ensuring that each of them is trained on the samples labelled by the other model.
This technique empirically improves the robustness of optimisation using pseudo-labelling in our experiments.
A brief explanation of the learning algorithm is as follows.
\begin{enumerate}
\item Sample two mini-batches $\mathcal{B}_f$ and $\mathcal{B}_g$. Pseudo-labels of $\mathcal{B}_f$ and $\mathcal{B}_g$ are given by $f$ and $g$, respectively.
\item Compute gradients of the loss functions and update $f$ on $\mathcal{B}_g$ and $g$ on $\mathcal{B}_f$. 
\end{enumerate}
Therefore, the loss functions for $f$ on $\mathcal{B}_g$ is defined as:
\begin{eqnarray}
  \mathcal{L}_{B}^{(f)} &=& \sum_{(u, i, l_{u,i}^{(g)}) \in \mathcal{B}_{g}}\ell\left(l_{u,i}^{(g)}, \hat{y}_{u,i}^{(f)};\Theta_f\right)
\end{eqnarray}
where $l_{u,i}^{(g)}$ is the label determined by $\hat{l}_{u}(i)$ in Eq.~(\ref{label-function}) for user $u$ and item $i$ based on the prediction of $g$ (see also Algorithm~\ref{alg:kapl}).
For training $g$, we also optimise $\mathcal{L}_{B}^{(g)}$ on $\mathcal{B}_f$.
Throughout this paper,
we utilise only $f$ as the final model to predict $\bf Y$ and do not choose or ensemble the models for a fair comparison of methods even there are two well-trained models after the training procedure.

\section{Experimental Settings}
\subsection{Baselines\label{baseline}}
We select state-of-the-art KG-based recommenders as the baselines, namely, RippleNet, MKR, KGAT, and KGNN-LS; besides, we introduce a naive baseline based on item popularity, TopPopular.  
\begin{itemize}[leftmargin=*]  
\item\textbf{RippleNet}~\cite{RippleNet}: This leverages multi-hop paths rooted at each user in a KG to enrich their representations and employs matrix factorisation on the representations.

\item\textbf{MKR}~\cite{wang2019multi}: This introduces a multi-task learning algorithm for recommendation and graph translation to enhance a recommender with translated graph embeddings.   
  
\item\textbf{KGAT}~\cite{KGAT}: This is a KG-based recommender, which employs GNN on a KG to generate the representations of users and items. This employs a pairwise loss function that optimises user wise recall directly. 

\item\textbf{KGNN-LS}~\cite{wang2019knowledge2}: This is also a GNN-based recommender, which employs a regularisation technique based on the label-smoothness assumption. This adopts a point-wise loss function that does not require extensive negative sampling.

\item\textbf{TopPopular}~\cite{dacrema2019we}: This is a non-personalised baseline that ranks items in the order of popularity, i.e., the number of users who have interacted with the item, in the train and validation splits.
  
\end{itemize}

Throughout our experiments, we conduct model selection for each method based on R@10 computed on the validation split.

\subsection{Datasets\label{dataset}}
We utilise three public datasets in our experiments;
(a) MovieLens1M\footnote{\url{https://grouplens.org/datasets/movielens/}}; (b) Last.FM\footnote{\url{https://grouplens.org/datasets/hetrec-2011/}}; and (c) BookCrossig\footnote{\url{http://www2.informatik.uni-freiburg.de/~cziegler/BX/}}.
KG data is available in public\footnote{https://github.com/hwwang55} and built upon Satori\footnote{https://searchengineland.com/library/bing/bing-satori}.

The statistics of the three datasets are listed in Table~\ref{table:statistics}.
The number of interactions in Table~\ref{table:statistics} indicates only those of the observed samples.
We observe that the three datasets have different levels of sparsity:
it is 2.66\%, 0.294\% and 0.026\% for MovieLens1M, Last.FM, and BookCrossing, respectively.

We split each dataset into training, validation and test splits at the ratio of 6:2:2.
For all datasets, we keep all the observed samples as implicit feedback.
Note that we created the negative samples for evaluation splits (i.e. validation and test splits)
while ensuring that those examples are unobserved in the entire dataset.
By contrast, because we consider implicit settings wherein we cannot observe true negative samples,
we randomly create negative samples for training
so that the negative samples are unobserved in the training and validation splits.
Thus, some negative samples for training can be positive in the test split.

\begin{table}[t]
  \centering
  \caption{Statistics of the datasets: MovieLens1M, Last.FM, and BookCrossing.}
  \begin{tabular}{c|c|c|c}
    \hline
    & MovieLens1M & Last.FM & BookCrossing\\
    \hline
    \# users        & 6,036   & 1,872     & 17,860  \\
    \# items        & 2,347   & 3,846     & 14,967  \\
    \# interactions & 376,886 & 21,173    & 69,873  \\
    \# sparsity     & 2.66\%  & 0.294\%   & 0.026\%  \\
    \# entities     & 102,569 & 9,366     & 77,903  \\
    \# relations    & 32      & 60        & 25      \\
    \# KG triples   & 499,474 & 15,518    & 151,500 \\
    \hline
  \end{tabular}
  \label{table:statistics}
\end{table}

\subsection{Implementation Details}
For our proposed KGPL, the number of convolutional layers $L$ is chosen from $\{1,2\}$,
and the dimensionality of latent space $d_l$ is 64 for all three datasets.
The sampled neighbour size is 32, 32, and 8 for MovieLens1M, Last.FM, and BookCrossing, respectively.
The depth of BFS, $h$, is 5, 6, and 6 for MovieLens1M, Last.FM, and BookCrossing, respectively.
The dropout ratio is tuned in ${0.1,0.2,\dots,0.8}$.
$a$ and $b$ are tuned in $\{10^{-3},10^{-2},0.1,\dots,1.0\}$ and $\{10^{-3},10^{-2},0.1,\dots,1.0\}$.
The learning rate is tuned in $\{1e-3,2e-3,\ldots,1e-2\}$, and the batch size is in $\{3333,6666,9999\}$.

\begin{table*}[h]
  \centering
    \caption{Results of $Precision@K$ and $Recall@K$ in top-$K$ recommendation.}
  \setlength{\tabcolsep}{1.5pt}
  \label{tbl:fulleval}
  \begin{tabular}{c|cccc|cccc|cccc|cccc|cccc|cccc}
    \hline
    Model & \multicolumn{8}{c|}{MovieLens1M} & \multicolumn{8}{c|}{Last.FM} & \multicolumn{8}{c}{BookCrossing} \\ \hline
    Measure
    & \multicolumn{4}{c|}{\textit{Precision@K}} & \multicolumn{4}{c|}{\textit{Recall@K}}
    & \multicolumn{4}{c|}{\textit{Precision@K}} & \multicolumn{4}{c|}{\textit{Recall@K}}
    & \multicolumn{4}{c|}{\textit{Precision@K}} & \multicolumn{4}{c}{\textit{Recall@K}} \\ \hline
    $K$
    & \textit{10} & \textit{20} & \textit{50} & \textit{100} & \textit{10} & \textit{20} & \textit{50} & \textit{100}
    & \textit{10} & \textit{20} & \textit{50} & \textit{100} & \textit{10} & \textit{20} & \textit{50} & \textit{100}
    & \textit{10} & \textit{20} & \textit{50} & \textit{100} & \textit{10} & \textit{20} & \textit{50} & \textit{100} \\ \hline
    TopPopular
    & .125 & .102 & .070 & .051 & .113 & .186 & .310 & .430 
    & .029 & .023 & .014 & .010 & .122 & .194 & .288 & .386 
    & \textbf{.017} & \textbf{.011} & \textbf{.006} & \textbf{.005} & \textbf{.083} & \textbf{.101} & \textbf{.143} & \textbf{.186} \\
    RippleNet~\cite{RippleNet}
    & .125 & .100 & .070 & .052 & .113 & .183 & .307 & .434  
    & .030 & .023 & .014 & .009 & .125 & .195 & .291 & .391  
    & .014 & .009 & .005 & .004 & .069 & .086 & .117 & .150 \\ 
    MKR~\cite{wang2019multi}
    & .123 & .100 & .069 & .051 & .110 & .182 & .303 & .425 
    & .029 & .022 & .014 & .010 & .121 & .181 & .293 & .397
    & .013 & .008 & .004 & .003 & .067 & .080 & .100 & .118 \\
    KGAT~\cite{KGAT}
    & \textbf{.188} & \textbf{.148} & \textbf{.100} & .069 & \textbf{.184} & .280 & .452 & .\textbf{603} 
    & .046 & .034 & .020 & .013 & .186 & .272 & .398 & .506
    & .008 & .006 & .004 & .003 & .044 & .059 & .093 & .132 \\
    KGNN-LS~\cite{wang2019knowledge2}
    & .139 & .113 & .080 & .058 & .121 & .195 & .335 & .475 
    & .029 & .022 & .014 & .009 & .122 & .182 & .289 & .386
    & .015 & .010 & .006 & .004 & .072 & .090 & .130 & .166 \\
    \hline
    KGPL
    & .177 & .144 & .099 & .069   & .183  & \textbf{.282} & \textbf{.454} & .602   
    & \textbf{.054} & \textbf{.039} & \textbf{.023} & \textbf{.014} & \textbf{.221} & \textbf{.314} & \textbf{.452} & \textbf{.557} 
    & .015 & .010 & .006 & .004 & .074 & .092 & .125 & .168 \\
    \hline
  \end{tabular}
\end{table*}

\begin{figure*}[h]
  \centering
  \includegraphics[clip,width=\linewidth]{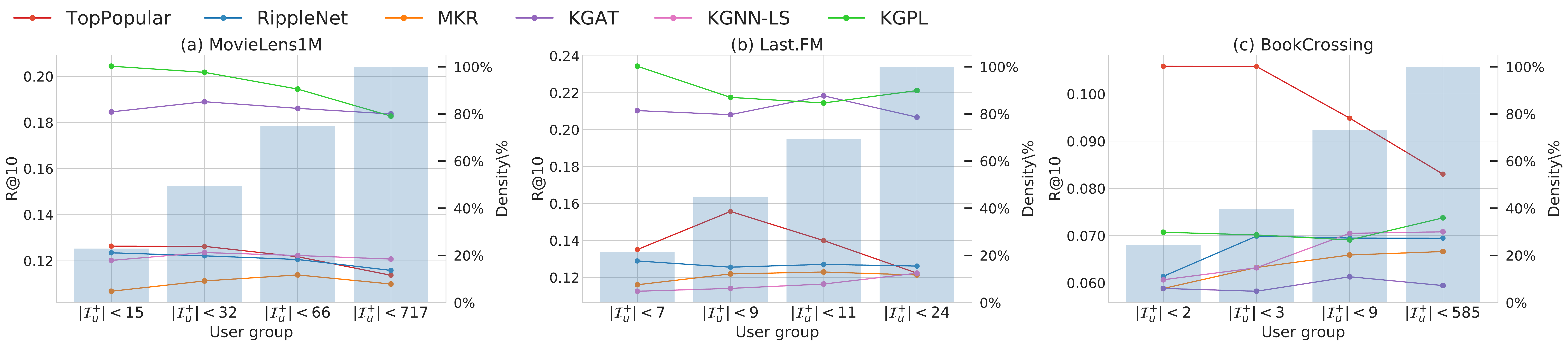}
  \caption{R@10 for Four User Groups with Different Sparsity Levels in (a) MovieLens1M, (b) Last.FM, and (c) BookCrossing.}
  \label{fig:user-group-stack}
\end{figure*}

\section{Experimental Results\label{result-section}}
\subsection{Overall Comparison with Baselines\label{main-result}}
We compare our method and the baseline methods
in terms of overall performance.
To compare the multiple recommenders,
we carefully conducted statistical significance testing through a paired Tukey HSD test with 95\% confidence intervals for each dataset in terms of R@10
and report the results also with effect sizes (Hedge's $g$) accordingly~\cite{brashlerCLEF01,sakai2018laboratory}.

Table~\ref{tbl:fulleval} lists the precision and recall ($k=10,20,50,100$) of each method on the three datasets.
Remarkably, on BookCrossing, TopPopular performs better than the other methods including ours, whereas only KGAT and MKR are statistically significantly worse than TopPopular ($p \leq2\mathrm{e}\mbox{-}16$, $g=0.104$ and $p=0.016$, $g=0.070$, respectively).
BookCrossing is an extremely sparse dataset,
and there are many users with only one observed interaction in the training split.
Thus, personalising recommendation may be challenging, and TopPopular is a strong baseline just by exploiting popularity bias.
We also find that RippleNet, MKR and KGNN-LS perform similarly to TopPopular on the other datasets.
We will investigate these observations more in Section~\ref{sec:item-side analysis}.

While KGPL outperforms the other methods on Last.FM,
KGPL performs slightly worse than KGAT for MovieLens1M; however, the difference between KGPL and KGAT is not statistically significant ($p=1.0$, $g=0.005$ and $p=0.635$, $g=0.047$ for MovieLens1M and Last.FM).
In BookCrossing, the performance of KGPL is rather limited.
This is because KGPL cannot take the benefits of pseudo-labelling on the extremely sparse dataset.

Overall, KGPL can consistently achieve comparative performance with the best model in each dataset.
Our baseline TopPopular also reveals a challenge in the personalised recommendation on severely sparse datasets.
It is also remarkable that KGPL shows significant improvement from KGNN-LS, which has almost the same neural network architecture as KGPL,
in MovieLens1M and Last.FM ($p\leq2\mathrm{e}\mbox{-}16$, $g=0.343$ and $p\leq2\mathrm{e}\mbox{-}16$, $g=0.354$).
This performance boost validates the effectiveness of our proposed method.

\subsection{Analysis on Users with Different Sparsity\label{user-group-result}}
For a more detailed comparison of the methods,
we created four groups of users with different sparsity levels according to the number of observed samples for each user in the training split
with thresholds of 25\%, 50\%, 75\%, and 100\% as in the previous work~\cite{KGAT}. 
Figure~\ref{fig:user-group-stack} (a)--(c) demonstrate R@10 of each method for each user group for the datasets of (a) MovieLens1M, (b) Last.FM and (c) BookCrossing.
Each bin corresponds to a group of users whose interactions is less than $x$. We set the threshold $x$ to 25\%, 50\%, 75\%, and 100\% percentile.
Therefore, the most left groups are cold-start users, and the most right ones show the performance on the whole users.
In Figure~\ref{fig:user-group-stack}~(a),
we can observe that KGAT and KGPL achieve consistently higher R@10 than the other methods for each user group.
Although KGPL performs slightly worse than KGAT in terms of overall performance ($|\mathcal{I}_u^{+}|<717$),
it achieves higher R@10 in the user group with the severe sparsity level ($|\mathcal{I}_u^{+}|<15$, $|\mathcal{I}_u^{+}|<32$, and $|\mathcal{I}_u^{+}|<66$).
In Figure~\ref{fig:user-group-stack}~(b),
KGPL performs best for three user groups ($|\mathcal{I}_u^{+}|<7$, $|\mathcal{I}_u^{+}|<9$, and $|\mathcal{I}_u^{+}|<24$),
and there is substantial improvement particularly for cold-start users ($|\mathcal{I}_u^{+}|<7$).
Figure~\ref{fig:user-group-stack}~(c) shows a similar trend.
It is remarkable that TopPopular shows extremely high performance for cold-start users and performs poorly for non-cold-start users in BookCrossing.
This result implies that the observed data for cold-start users (i.e. less frequent users in the observed dataset) is biased toward popularity. 

As a summary,
KGPL can successfully adapt to various sparsity levels of datasets and shows improvements in performance, particularly for cold-start users.
The improvement of KGPL is relatively large for Last.FM, whereas those for MovieLens1M and BookCrossing are mild.
The results suggest the advantage and limitation of KGPL;
it can prove its merits for real-world datasets, yet the gain is rather mild for dense or highly sparse datasets (e.g. MovieLens1M and BookCrossing) in terms of overall performance.

\section{In-Depth Analyses}
As our main concern is the performance for cold-start users or items,
we investigate the behaviours of the baseline recommenders and KGPL for users and items with different frequencies (i.e. the number of observed data) from different angles in this section. 

\subsection{User-Side Performance Analysis\label{win-analysis}}
We analyse how many users can benefit from each recommender
rather than the overall average of R@10.
To this end, for each user in the test split,
we record a \emph{winner system} which achieves the highest R@10.
If there are ties, we choose all of the tied systems as winners
while we do not choose any systems when the R@10 of all systems is 0.0.
Then, for each system, we count the number of wins for all users.
As we hypothesised that our pseudo-labelling is effective for alleviating user-side cold-start problems,
we also consider a variant of KGPL, $\mbox{KGPL}_{nopl}$, which does not augment labels through pseudo-labelling.

\begin{figure*}[t!]
    \centering
    \includegraphics[clip,width=0.9\linewidth]{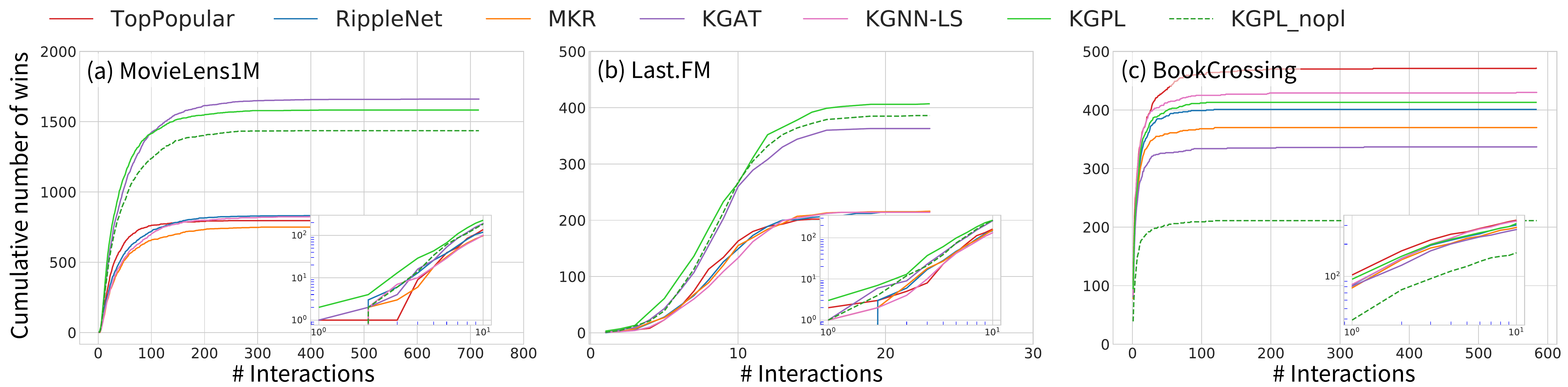}
    \caption{Cumulative number of wins for users with different number of observed interactions.
     }
    \label{fig:freq-vs-win}
\end{figure*}

Figure~\ref{fig:freq-vs-win} shows the number of wins for each method 
when the upper bound of the number of observed interactions for a user is varied.  
The $x$- and $y$-axes in the figure indicate the number of observed interactions in the training split
and the cumulative number of wins for the set of users with, at most, $x$ interactions, respectively.
The inset shown in each part shows the result for users with less than 10 interactions,
and the $y$-axis is shown with a logarithmic scale.
In Figure~\ref{fig:freq-vs-win} (a),
KGAT and KGPL outperform the other methods as in the results of Section~\ref{result-section},
while KGPL can achieve better performance for more cold-start users than KGAT (see the small figure in Figure~\ref{fig:freq-vs-win} (a)).
Figure~\ref{fig:freq-vs-win} (b) also demonstrates the advantage of KGPL;
in addition to cold-start users, KGPL shows substantial improvement also in users with more than ten interactions.
This result is intuitive because KGPL can augment pseudo-positive samples more precisely for heavy users.
The comparison between KGPL and $\mbox{KGPL}_{nopl}$ in MovieLens1M and Last.FM demonstrates
the effectiveness of our pseudo-labelling approach.
In BookCrossing, KGPL achieves slightly better performance particularly for cold-start users than the other KG-aware methods,
whereas TopPopular performs best among all of the methods.

Based on these results, we conclude that
our KGPL can provide better recommendation results for a wide range of users through the proposed pseudo-labelling approach.
However, the results on BookCrossing reveals that developing a recommender for extremely sparse datasets is still challenging even with KGs.

\subsection{Item-Side Performance Analysis}
\label{sec:item-side analysis}
The effectiveness of the popularity-based method (i.e. TopPopular) indicates that the popularity of items is a strong clue in recommendation problems.
Nevertheless, a method heavily relying on popularity may impede the collection of informative users' feedback
by narrowing the coverage of exposed items.
Therefore, we examine the coverage of items in the ranked lists by each method.

\begin{figure*}[htpb]
    \centering
    \includegraphics[clip,width=\linewidth]{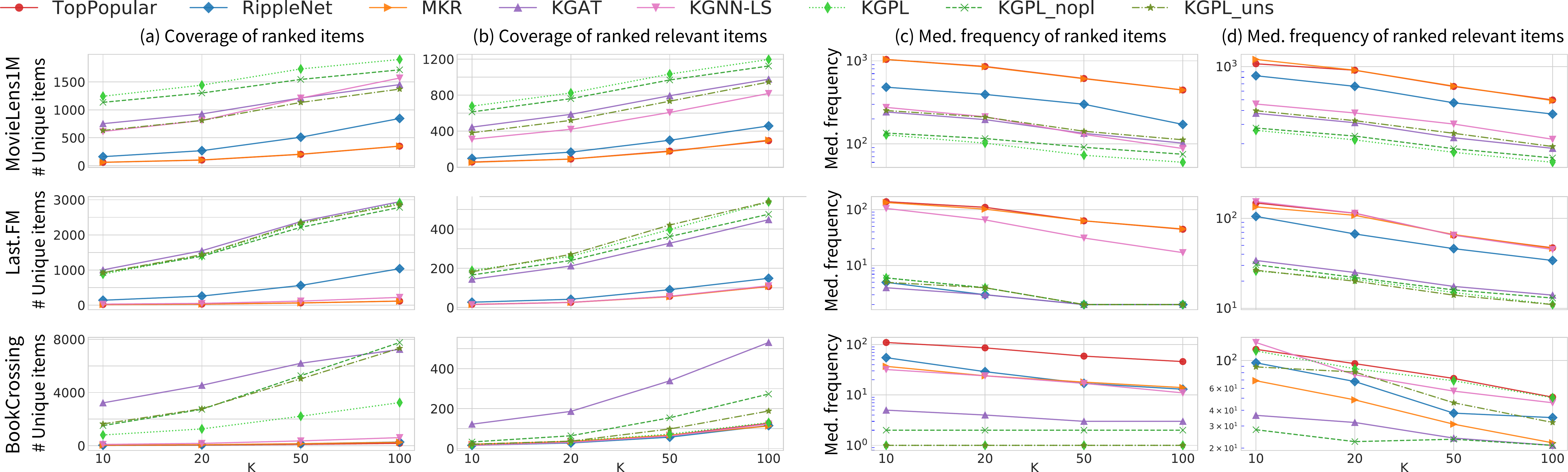}
    \caption{Analysis of item coverage in the ranked lists.
      Each row from the top shows the results on MovieLens1M, Last.FM and BookCrossing, respectively.}
    \label{fig:item-coverage}
\end{figure*}

To this end, we calculate the number of unique items ranked at least once in a top-$K$ ranking for a user as relevant items.
We also check the number of unique items in the top-$K$ rankings regardless of relevance.
Figure~\ref{fig:item-coverage} shows the results in the three datasets,
and each row from the top is corresponding to MovieLens1M, Last.FM, and BookCrossing, respectively.
Columns from the left side in each row indicate
(a) the number of ranked unique items;
(b) the number of unique items that are ranked as relevant at least once;
(c) the median frequency of ranked unique items; and
(d) the median frequency of relevant items.
A large number of unique items in the ranked lists indicates that the method provides highly personalised results.
TopPopular shows the lowest coverage and highest median frequency of items as it provides identical results for all users based on popularity.
To examine the proposed negative sampling strategy, we test another variant of KGPL, $\mbox{KGPL}_{uns}$, which samples negative instances according to a uniform distribution.

In MovieLens1M,
the coverage of items of KGPL is higher than those of the other methods with/without regard to the relevance of items (see the first row and columns (a) and (b)).
KGPL also show lower values for the median frequency of ranked items (the first row in columns (c) and (d)).
This suggests that KGPL tries to pull up less popular but relevant items. 
In Last.FM (the second row of Figure~\ref{fig:item-coverage}),
KGAT and KGPL show a similar coverage of ranked items (column (a));
however, KGPL achieves clearly a higher coverage for relevant items than that of KGAT (column (b)).
In terms of the median frequency of items, KGPL exhibits a lower median frequency of the relevant items as in MovieLens1M.
These results indicate that KGPL successfully adjust recommendation results for users,
whereas it slightly underperforms KGAT in MovieLens1M (discussed in Section~\ref{result-section}).

The results of BookCrossing show a different trend.
Based on columns (a) and (b) in the third row, KGAT shows the highest coverage,
and the median frequency of ranked relevant items is consistently smaller than those of the other methods.
However, it should be noted that KGAT compromises the overall accuracy in BookCrossing (Section~\ref{result-section})
and ranks nonrelevant items more in its top-K results.
KGPL shows a higher coverage of relevant/nonrelevant items than RippleNet, MKR, and KGNN-LS (column (a)),
but for relevant items, the coverage of KGPL is almost the same as those of the other three methods.
These results imply that KGPL captures the popularity bias in BookCrossing
and compromises the coverage of items.
This result is reasonable because the relevant items in BookCrossing are extremely biased by their popularity. 

Comparison of KGPL and its variants show some positive effects of the popularity-aware negative sampling and pseudo-labelling on the coverage of items. By comparing $\mbox{KGPL}_{nopl}$ and KGPL, we observe that pseudo-labelling improves the coverage of items in MovieLens1M and Last.fm.
In MovieLens1M, $\mbox{KGPL}_{uns}$ shows a drop in coverage.
Without adjusting the distribution for negative sampling, KGPL relies on popularity biases and thus loses the personalised results.
For BookCrossing, due to its extreme sparsity, KGPL and its variants fail to personalise the recommendation results.
BookCrossing also has a strong popularity bias; thus the diversity of items and the accuracy of recommendation are in a relation of a trade-off to some extent. This prevents KGPL from improving the coverage of items. 

For BookCrossing and Last.FM,
the item coverage of RippleNet, MKR and KGNN-L is relatively low (see columns (a) and (b)),
and the median frequency of items is high.
In particular, for Last.FM,
these three methods show extremely low coverage and high median frequency of relevant items,
and the overall performance of these methods is similar to that of TopPopular (Section~\ref{result-section}).
These results imply these methods collapse to TopPopular and excessively exploit the popularity bias when a dataset is sparse.

\section{Ablation Analysis}
KGPL comprises three main components:
(1) mini-batch augmentation through pseudo-labelling;
(2) knowledge-aware item sampling for pseudo-labelling; and
(3) co-training approach for stable training.
To demonstrate the effectiveness of each component,
we conduct ablation studies by comparing KGPL with its variants:
\begin{itemize}
\item $\mbox{KGPL}_{nopl}$: KGPL without the pseudo-labelling.
\item $\mbox{KGPL}_{rand,self}$: KGPL without the knowledge-aware sampling strategy and co-training technique.
\item $\mbox{KGPL}_{kapl,self}$: KGPL without the co-training technique.
\item $\mbox{KGPL}_{rand,cot}$: KGPL without the knowledge-aware sampling.
\item $\mbox{KGPL}_{kapl,cot}$: The full model of KGPL. 
\end{itemize}
Figure~\ref{fig:curve_ablation} shows the curve of the validation R@10 of each KGPL variants in the training process in the Last.FM dataset.
Table~\ref{table:ablation_table} lists the precision and recall results for $k=10,20,50,100$ of each KGPL variants in the Last.FM dataset.

\subsection{Effects of the Co-Training}
\begin{figure}[htpb]
    \centering
    \includegraphics[height=4.0cm]{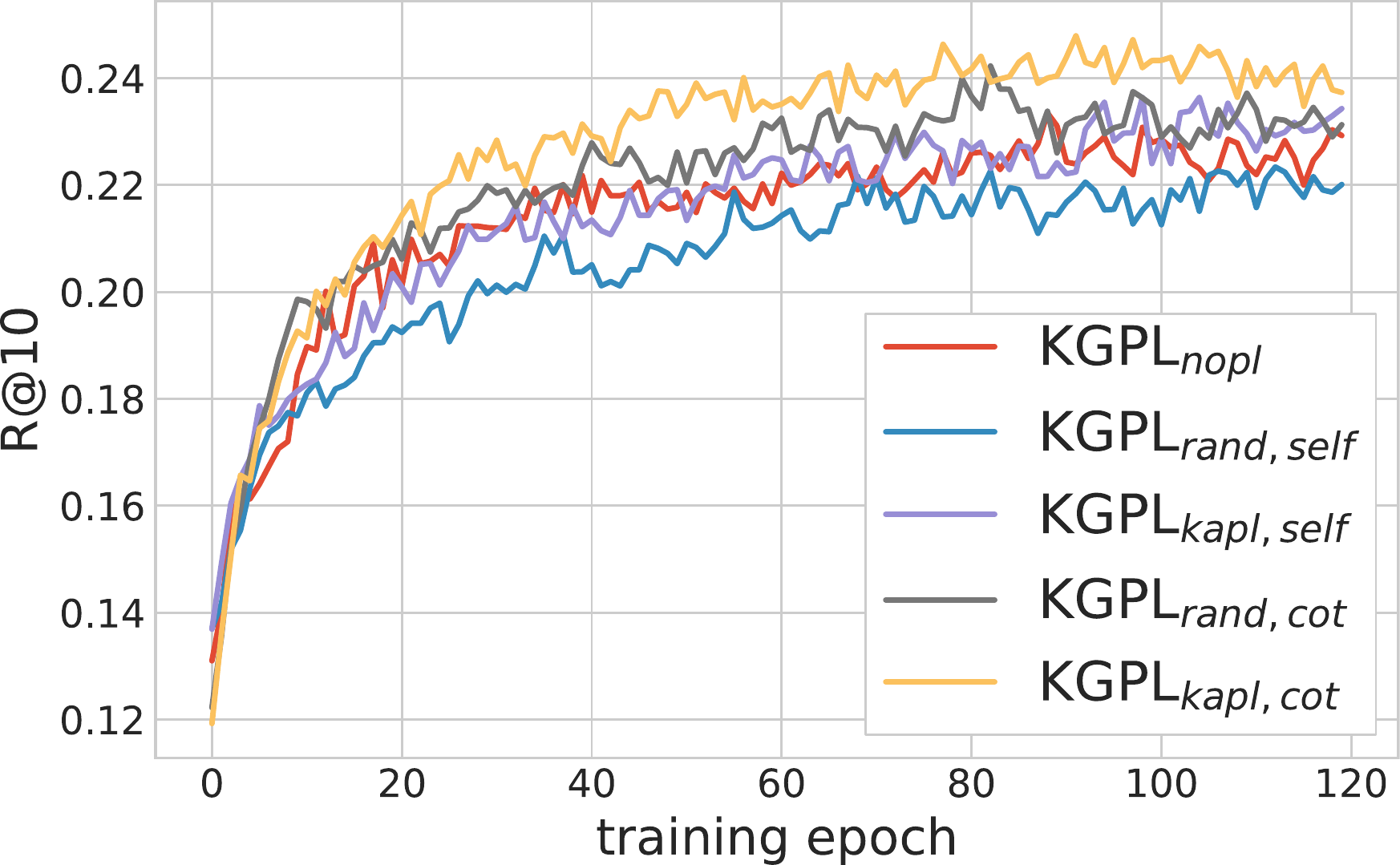}
    \caption{Training process of KGPL and its variants.}
    \label{fig:curve_ablation}
\end{figure}

Comparison between $\mbox{KGPL}_{rand,self}$ (blue line in Figure~\ref{fig:curve_ablation}) and $\mbox{KGPL}_{rand,cot}$ (grey line) demonstrate that only the co-training technique achieves substantial performance gain.
The same trend can be observed between $\mbox{KGPL}_{kapl,self}$ (purple line) and $\mbox{KGPL}_{kapl,cot}$ (yellow line).
Therefore, the co-training technique consistently improves performance.

\subsection{Effects of the Knowledge-Aware Item Sampling}
Pseudo-labelling without knowledge-aware sampling $\mbox{KGPL}_{rand,self}$ (blue line) performs poorly and even worse than the model without pseudo-labelling $\mbox{KGPL}_{nopl}$ (red line).
Solely introducing knowledge-aware sampling does not make full use of pseudo-labelling.
$\mbox{KGPL}_{kapl,self}$ (purple line) still does not outperform $\mbox{KGPL}_{nopl}$.
By contrast, when the co-training technique is used, knowledge-aware sampling provides a substantial performance lift as $\mbox{KGPL}_{rand,cot}$ (grey line) and $\mbox{KGPL}_{kapl,cot}$ (yellow line) show.
The knowledge-aware sampling enhances the stability of the training on pseudo-labels, and the performance gain is substantial particularly when used with the co-training approach. 

\begin{table}[t!]
  \centering
  \caption{Comparison of KGPL and its ablated variants.}
\setlength{\tabcolsep}{2pt}
\small
  \begin{tabular}{c|cccc|cccc}
    \hline
    Measure & \multicolumn{4}{c}{\textit{Precision@K}} & \multicolumn{4}{c}{\textit{Recall@K}} \\\hline
    $K$      & \textit{10} & \textit{20} & \textit{50} & \textit{100} & \textit{10} & \textit{20} & \textit{50} & \textit{100} \\
    \hline
    $\mbox{KGPL}_{nopl}$
    & .048 & .033 & .016 & .006 & .209 & .304 & .437 & .522 \\
    $\mbox{KGPL}_{rand,self}$
    & .050 & .037 & .022 & .013 & .204 & .298 & .436 & .531 \\ 
    $\mbox{KGPL}_{kapl,self}$
    & .051 & .037 & .022 & .013 & .208 & .303 & .436 & .537 \\
    $\mbox{KGPL}_{rand,cot}$
    & .052 & .036 & .018 & .009 & .216 & .317 & .448 & .545  \\  
    $\mbox{KGPL}_{kapl,cot}$
    & .054 & .039 & .023 & .014 & .221 & .314 & .452 & .557 \\
    \hline
  \end{tabular}
  \label{table:ablation_table}
\end{table}

\subsection{Hyper-Parameter Sensitivity}
\begin{figure}[htpb]
    \centering
    \includegraphics[height=2.7cm]{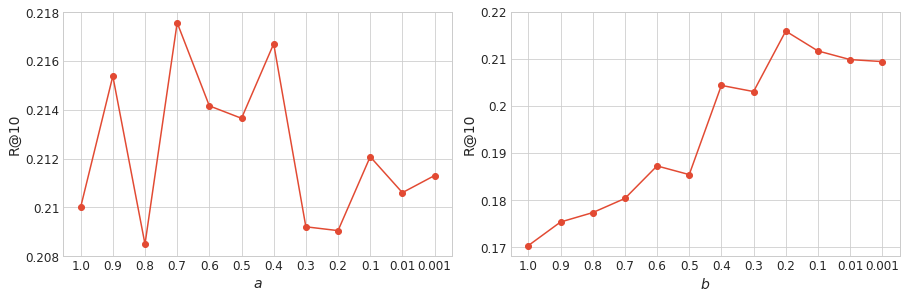}
    \caption{Impact of hyper-parameter $a$ and $b$.}
    \label{fig:hyperparam-analysis}
\end{figure}

KGPL has two hyper-parameters $a$ and $b$, which
control the skewness of sampling distributions for pseudo-labelling and negative sampling, respectively.
Figure~\ref{fig:hyperparam-analysis} shows the effect of $a$ and $b$ on R@10.
In the top figure, we observe that KGPL is highly stable for the value of $a$;
In the range from 0.7 to 0.4, KGPL consistently achieves high R@10.
The result suggests that both exploitation and exploration are important in the sampling of items for pseudo-labelling.
The figure shows that KGPL deteriorates substantially with a large $b$.
Sampling negatives with a large $b$ chooses popular items more as negatives,
and can impede the exploitation of the popularity bias.

\section{Conclusion}
We have proposed a KG-aware recommender, KGPL, for alleviating cold-start problems for both users and items.
Our method incorporates pseudo-labelling 
to explicitly leverage unobserved samples as weak positive/negative instances.
To enhance the training procedure with pseudo-labelled samples,
we have developed a KG-aware sampling strategy and a co-training approach in which two different models provide supervisions with each other.

The experimental results demonstrate that our KGPL achieves performance comparable with state-of-the-art KG-aware recommenders in the multiple datasets.
KGPL outperforms the other methods, particularly for cold-start users and items.
Our extensive analyses indicate that KGPL provides more personalised recommendation results and discovers relevant items more from cold-start items than the baselines without compromising overall performance.

As future work,
we would like to explore an adaptive sampling for pseudo-labelling such as that presented by Wang~\emph{et al.}~\cite{KGPolicy20}.
It would also be interesting to investigate the connectivity between GNNs with many convolutional layers
and our proposed pseudo-labelling approach;
our method may help to propagate gradients from observed samples to unobserved ones, and therefore examining architectures such as residual networks may be insightful for solving cold-start problems by GNNs.
Combining our approach and recent non-sampling learning methods~\cite{chen2020jointly} is also interesting; our approach can provide adaptive weights for each user-item entry in their loss function with model predictions.

%% file: head.bbl

\begin{thebibliography}{46}


\ifx \showCODEN    \undefined \def \showCODEN     #1{\unskip}     \fi
\ifx \showDOI      \undefined \def \showDOI       #1{#1}\fi
\ifx \showISBNx    \undefined \def \showISBNx     #1{\unskip}     \fi
\ifx \showISBNxiii \undefined \def \showISBNxiii  #1{\unskip}     \fi
\ifx \showISSN     \undefined \def \showISSN      #1{\unskip}     \fi
\ifx \showLCCN     \undefined \def \showLCCN      #1{\unskip}     \fi
\ifx \shownote     \undefined \def \shownote      #1{#1}          \fi
\ifx \showarticletitle \undefined \def \showarticletitle #1{#1}   \fi
\ifx \showURL      \undefined \def \showURL       {\relax}        \fi
\providecommand\bibfield[2]{#2}
\providecommand\bibinfo[2]{#2}
\providecommand\natexlab[1]{#1}
\providecommand\showeprint[2][]{arXiv:#2}

\bibitem[\protect\citeauthoryear{Berg, Kipf, and Welling}{Berg
  et~al\mbox{.}}{2017}]%
        {berg2017graph}
\bibfield{author}{\bibinfo{person}{Rianne van~den Berg},
  \bibinfo{person}{Thomas~N Kipf}, {and} \bibinfo{person}{Max Welling}.}
  \bibinfo{year}{2017}\natexlab{}.
\newblock \showarticletitle{Graph convolutional matrix completion}.
\newblock \bibinfo{journal}{\emph{arXiv preprint arXiv:1706.02263}}
  (\bibinfo{year}{2017}).
\newblock


\bibitem[\protect\citeauthoryear{Blum and Mitchell}{Blum and Mitchell}{1998}]%
        {blum1998combining}
\bibfield{author}{\bibinfo{person}{Avrim Blum} {and} \bibinfo{person}{Tom
  Mitchell}.} \bibinfo{year}{1998}\natexlab{}.
\newblock \showarticletitle{Combining labeled and unlabeled data with
  co-training}. In \bibinfo{booktitle}{\emph{Proceedings of the Eleventh Annual
  Conference on Computational Learning Theory}}. \bibinfo{pages}{92--100}.
\newblock


\bibitem[\protect\citeauthoryear{Braschler}{Braschler}{2001}]%
        {brashlerCLEF01}
\bibfield{author}{\bibinfo{person}{Martin Braschler}.}
  \bibinfo{year}{2001}\natexlab{}.
\newblock \showarticletitle{CLEF 2001 - Overview of Results}. In
  \bibinfo{booktitle}{\emph{Revised Papers from the Second Workshop of the
  Cross-Language Evaluation Forum on Evaluation of Cross-Language Information
  Retrieval Systems}}. \bibinfo{pages}{9–26}.
\newblock


\bibitem[\protect\citeauthoryear{Bruna, Zaremba, Szlam, and LeCun}{Bruna
  et~al\mbox{.}}{2013}]%
        {bruna2013spectral}
\bibfield{author}{\bibinfo{person}{Joan Bruna}, \bibinfo{person}{Wojciech
  Zaremba}, \bibinfo{person}{Arthur Szlam}, {and} \bibinfo{person}{Yann
  LeCun}.} \bibinfo{year}{2013}\natexlab{}.
\newblock \showarticletitle{Spectral networks and locally connected networks on
  graphs}.
\newblock \bibinfo{journal}{\emph{arXiv preprint arXiv:1312.6203}}
  (\bibinfo{year}{2013}).
\newblock


\bibitem[\protect\citeauthoryear{Chen, Zhang, Ma, Liu, and Ma}{Chen
  et~al\mbox{.}}{2020}]%
        {chen2020jointly}
\bibfield{author}{\bibinfo{person}{Chong Chen}, \bibinfo{person}{Min Zhang},
  \bibinfo{person}{Weizhi Ma}, \bibinfo{person}{Yiqun Liu}, {and}
  \bibinfo{person}{Shaoping Ma}.} \bibinfo{year}{2020}\natexlab{}.
\newblock \showarticletitle{Jointly Non-Sampling Learning for Knowledge Graph
  Enhanced Recommendation}. In \bibinfo{booktitle}{\emph{Proceedings of
  SIGIR}}.
\newblock


\bibitem[\protect\citeauthoryear{Dacrema, Cremonesi, and Jannach}{Dacrema
  et~al\mbox{.}}{2019}]%
        {dacrema2019we}
\bibfield{author}{\bibinfo{person}{Maurizio~Ferrari Dacrema},
  \bibinfo{person}{Paolo Cremonesi}, {and} \bibinfo{person}{Dietmar Jannach}.}
  \bibinfo{year}{2019}\natexlab{}.
\newblock \showarticletitle{Are we really making much progress? A worrying
  analysis of recent neural recommendation approaches}. In
  \bibinfo{booktitle}{\emph{Proceedings of the 13th ACM Conference on
  Recommender Systems}}. \bibinfo{pages}{101--109}.
\newblock


\bibitem[\protect\citeauthoryear{Defferrard, Bresson, and
  Vandergheynst}{Defferrard et~al\mbox{.}}{2016}]%
        {defferrard2016convolutional}
\bibfield{author}{\bibinfo{person}{Micha{\"e}l Defferrard},
  \bibinfo{person}{Xavier Bresson}, {and} \bibinfo{person}{Pierre
  Vandergheynst}.} \bibinfo{year}{2016}\natexlab{}.
\newblock \showarticletitle{Convolutional neural networks on graphs with fast
  localized spectral filtering}. In \bibinfo{booktitle}{\emph{Advances in
  Neural Information Processing Systems}}. \bibinfo{pages}{3844--3852}.
\newblock


\bibitem[\protect\citeauthoryear{Douze, Szlam, Hariharan, and Jégou}{Douze
  et~al\mbox{.}}{2018}]%
        {Douze_2018_CVPR}
\bibfield{author}{\bibinfo{person}{Matthijs Douze}, \bibinfo{person}{Arthur
  Szlam}, \bibinfo{person}{Bharath Hariharan}, {and} \bibinfo{person}{Hervé
  Jégou}.} \bibinfo{year}{2018}\natexlab{}.
\newblock \showarticletitle{Low-Shot Learning With Large-Scale Diffusion}. In
  \bibinfo{booktitle}{\emph{Proceedings of the IEEE Conference on Computer
  Vision and Pattern Recognition (CVPR)}}.
\newblock


\bibitem[\protect\citeauthoryear{Duvenaud, Maclaurin, Iparraguirre, Bombarell,
  Hirzel, Aspuru-Guzik, and Adams}{Duvenaud et~al\mbox{.}}{2015}]%
        {duvenaud2015convolutional}
\bibfield{author}{\bibinfo{person}{David~K Duvenaud}, \bibinfo{person}{Dougal
  Maclaurin}, \bibinfo{person}{Jorge Iparraguirre}, \bibinfo{person}{Rafael
  Bombarell}, \bibinfo{person}{Timothy Hirzel}, \bibinfo{person}{Al{\'a}n
  Aspuru-Guzik}, {and} \bibinfo{person}{Ryan~P Adams}.}
  \bibinfo{year}{2015}\natexlab{}.
\newblock \showarticletitle{Convolutional networks on graphs for learning
  molecular fingerprints}. In \bibinfo{booktitle}{\emph{Advances in neural
  information processing systems}}. \bibinfo{pages}{2224--2232}.
\newblock


\bibitem[\protect\citeauthoryear{Fu, Xian, Gao, Zhao, Huang, Ge, Xu, Geng,
  Shah, Zhang, and de~Melo}{Fu et~al\mbox{.}}{2020}]%
        {fairnessZuohui20}
\bibfield{author}{\bibinfo{person}{Zuohui Fu}, \bibinfo{person}{Yikun Xian},
  \bibinfo{person}{Ruoyuan Gao}, \bibinfo{person}{Jieyu Zhao},
  \bibinfo{person}{Qiaoying Huang}, \bibinfo{person}{Yingqiang Ge},
  \bibinfo{person}{Shuyuan Xu}, \bibinfo{person}{Shijie Geng},
  \bibinfo{person}{Chirag Shah}, \bibinfo{person}{Yongfeng Zhang}, {and}
  \bibinfo{person}{Gerard de Melo}.} \bibinfo{year}{2020}\natexlab{}.
\newblock \showarticletitle{Fairness-Aware Explainable Recommendation over
  Knowledge Graphs}. In \bibinfo{booktitle}{\emph{Proceedings of the 43rd
  International ACM SIGIR Conference on Research and Development in Information
  Retrieval}}. \bibinfo{pages}{69–78}.
\newblock


\bibitem[\protect\citeauthoryear{Hamilton, Ying, and Leskovec}{Hamilton
  et~al\mbox{.}}{2017}]%
        {hamilton2017inductive}
\bibfield{author}{\bibinfo{person}{Will Hamilton}, \bibinfo{person}{Zhitao
  Ying}, {and} \bibinfo{person}{Jure Leskovec}.}
  \bibinfo{year}{2017}\natexlab{}.
\newblock \showarticletitle{Inductive representation learning on large graphs}.
  In \bibinfo{booktitle}{\emph{Advances in Neural Information Processing
  Systems}}. \bibinfo{pages}{1024--1034}.
\newblock


\bibitem[\protect\citeauthoryear{Han, Yao, Yu, Niu, Xu, Hu, Tsang, and
  Sugiyama}{Han et~al\mbox{.}}{2018}]%
        {han2018co}
\bibfield{author}{\bibinfo{person}{Bo Han}, \bibinfo{person}{Quanming Yao},
  \bibinfo{person}{Xingrui Yu}, \bibinfo{person}{Gang Niu},
  \bibinfo{person}{Miao Xu}, \bibinfo{person}{Weihua Hu}, \bibinfo{person}{Ivor
  Tsang}, {and} \bibinfo{person}{Masashi Sugiyama}.}
  \bibinfo{year}{2018}\natexlab{}.
\newblock \showarticletitle{Co-teaching: Robust training of deep neural
  networks with extremely noisy labels}. In \bibinfo{booktitle}{\emph{Advances
  in Neural Information Processing Systems}}. \bibinfo{pages}{8527--8537}.
\newblock


\bibitem[\protect\citeauthoryear{He, Zhang, Kan, and Chua}{He
  et~al\mbox{.}}{2016}]%
        {he2016fast}
\bibfield{author}{\bibinfo{person}{Xiangnan He}, \bibinfo{person}{Hanwang
  Zhang}, \bibinfo{person}{Min-Yen Kan}, {and} \bibinfo{person}{Tat-Seng
  Chua}.} \bibinfo{year}{2016}\natexlab{}.
\newblock \showarticletitle{Fast matrix factorization for online recommendation
  with implicit feedback}. In \bibinfo{booktitle}{\emph{Proceedings of the 39th
  International ACM SIGIR conference on Research and Development in Information
  Retrieval}}. \bibinfo{pages}{549--558}.
\newblock


\bibitem[\protect\citeauthoryear{Hu, Shi, Zhao, and Yu}{Hu
  et~al\mbox{.}}{2018a}]%
        {MCRec}
\bibfield{author}{\bibinfo{person}{Binbin Hu}, \bibinfo{person}{Chuan Shi},
  \bibinfo{person}{Wayne~Xin Zhao}, {and} \bibinfo{person}{Philip~S. Yu}.}
  \bibinfo{year}{2018}\natexlab{a}.
\newblock \showarticletitle{Leveraging Meta-path based Context for Top- {N}
  Recommendation with {A} Neural Co-Attention Model}. In
  \bibinfo{booktitle}{\emph{{Proceedings of the 24th ACM SIGKDD International
  Conference on Knowledge Discovery \& Data Mining}}}.
  \bibinfo{pages}{1531--1540}.
\newblock


\bibitem[\protect\citeauthoryear{Hu, Shi, Zhao, and Yu}{Hu
  et~al\mbox{.}}{2018b}]%
        {hu2018leveraging}
\bibfield{author}{\bibinfo{person}{Binbin Hu}, \bibinfo{person}{Chuan Shi},
  \bibinfo{person}{Wayne~Xin Zhao}, {and} \bibinfo{person}{Philip~S Yu}.}
  \bibinfo{year}{2018}\natexlab{b}.
\newblock \showarticletitle{Leveraging meta-path based context for top-n
  recommendation with a neural co-attention model}. In
  \bibinfo{booktitle}{\emph{Proceedings of the 24th ACM SIGKDD International
  Conference on Knowledge Discovery \& Data Mining}}.
  \bibinfo{pages}{1531--1540}.
\newblock


\bibitem[\protect\citeauthoryear{Hu, Koren, and Volinsky}{Hu
  et~al\mbox{.}}{2008}]%
        {hu2008collaborative}
\bibfield{author}{\bibinfo{person}{Yifan Hu}, \bibinfo{person}{Yehuda Koren},
  {and} \bibinfo{person}{Chris Volinsky}.} \bibinfo{year}{2008}\natexlab{}.
\newblock \showarticletitle{Collaborative filtering for implicit feedback
  datasets}. In \bibinfo{booktitle}{\emph{2008 Eighth IEEE International
  Conference on Data Mining}}. \bibinfo{pages}{263--272}.
\newblock


\bibitem[\protect\citeauthoryear{Huang, Zhao, Dou, Wen, and Chang}{Huang
  et~al\mbox{.}}{2018}]%
        {huang2018improving}
\bibfield{author}{\bibinfo{person}{Jin Huang}, \bibinfo{person}{Wayne~Xin
  Zhao}, \bibinfo{person}{Hongjian Dou}, \bibinfo{person}{Ji-Rong Wen}, {and}
  \bibinfo{person}{Edward~Y Chang}.} \bibinfo{year}{2018}\natexlab{}.
\newblock \showarticletitle{Improving sequential recommendation with
  knowledge-enhanced memory networks}. In \bibinfo{booktitle}{\emph{The 41st
  International ACM SIGIR Conference on Research \& Development in Information
  Retrieval}}. \bibinfo{pages}{505--514}.
\newblock


\bibitem[\protect\citeauthoryear{Iscen, Tolias, Avrithis, and Chum}{Iscen
  et~al\mbox{.}}{2019}]%
        {iscen2019label}
\bibfield{author}{\bibinfo{person}{Ahmet Iscen}, \bibinfo{person}{Giorgos
  Tolias}, \bibinfo{person}{Yannis Avrithis}, {and} \bibinfo{person}{Ondrej
  Chum}.} \bibinfo{year}{2019}\natexlab{}.
\newblock \showarticletitle{Label propagation for deep semi-supervised
  learning}. In \bibinfo{booktitle}{\emph{Proceedings of the IEEE conference on
  computer vision and pattern recognition}}. \bibinfo{pages}{5070--5079}.
\newblock


\bibitem[\protect\citeauthoryear{Karasuyama and Mamitsuka}{Karasuyama and
  Mamitsuka}{2013}]%
        {karasuyama2013manifold}
\bibfield{author}{\bibinfo{person}{Masayuki Karasuyama} {and}
  \bibinfo{person}{Hiroshi Mamitsuka}.} \bibinfo{year}{2013}\natexlab{}.
\newblock \showarticletitle{Manifold-based similarity adaptation for label
  propagation}. In \bibinfo{booktitle}{\emph{Advances in Neural Information
  Processing Systems}}. \bibinfo{pages}{1547--1555}.
\newblock


\bibitem[\protect\citeauthoryear{Kipf and Welling}{Kipf and Welling}{2016}]%
        {kipf2016semi}
\bibfield{author}{\bibinfo{person}{Thomas~N Kipf} {and} \bibinfo{person}{Max
  Welling}.} \bibinfo{year}{2016}\natexlab{}.
\newblock \showarticletitle{Semi-supervised classification with graph
  convolutional networks}.
\newblock \bibinfo{journal}{\emph{arXiv preprint arXiv:1609.02907}}
  (\bibinfo{year}{2016}).
\newblock


\bibitem[\protect\citeauthoryear{Lee}{Lee}{2013}]%
        {lee2013pseudo}
\bibfield{author}{\bibinfo{person}{Dong-Hyun Lee}.}
  \bibinfo{year}{2013}\natexlab{}.
\newblock \showarticletitle{Pseudo-label: The simple and efficient
  semi-supervised learning method for deep neural networks}. In
  \bibinfo{booktitle}{\emph{Workshop On Challenges In Representation Learning,
  ICML}}, Vol.~\bibinfo{volume}{3}.
\newblock


\bibitem[\protect\citeauthoryear{Lu, Fang, and Shi}{Lu et~al\mbox{.}}{2020}]%
        {lu2020meta}
\bibfield{author}{\bibinfo{person}{Yuanfu Lu}, \bibinfo{person}{Yuan Fang},
  {and} \bibinfo{person}{Chuan Shi}.} \bibinfo{year}{2020}\natexlab{}.
\newblock \showarticletitle{Meta-learning on heterogeneous information networks
  for cold-start recommendation}.
\newblock  (\bibinfo{year}{2020}).
\newblock


\bibitem[\protect\citeauthoryear{Maas, Hannun, and Ng}{Maas
  et~al\mbox{.}}{2013}]%
        {maas2013rectifier}
\bibfield{author}{\bibinfo{person}{Andrew~L Maas}, \bibinfo{person}{Awni~Y
  Hannun}, {and} \bibinfo{person}{Andrew~Y Ng}.}
  \bibinfo{year}{2013}\natexlab{}.
\newblock \showarticletitle{Rectifier nonlinearities improve neural network
  acoustic models}. In \bibinfo{booktitle}{\emph{ICML Workshop On Deep Learning
  For Audio, Speech And Language Processing}}.
\newblock


\bibitem[\protect\citeauthoryear{Monti, Boscaini, Masci, Rodola, Svoboda, and
  Bronstein}{Monti et~al\mbox{.}}{2017}]%
        {monti2017geometric}
\bibfield{author}{\bibinfo{person}{Federico Monti}, \bibinfo{person}{Davide
  Boscaini}, \bibinfo{person}{Jonathan Masci}, \bibinfo{person}{Emanuele
  Rodola}, \bibinfo{person}{Jan Svoboda}, {and} \bibinfo{person}{Michael~M
  Bronstein}.} \bibinfo{year}{2017}\natexlab{}.
\newblock \showarticletitle{Geometric deep learning on graphs and manifolds
  using mixture model cnns}. In \bibinfo{booktitle}{\emph{Proceedings of the
  IEEE Conference on Computer Vision and Pattern Recognition}}.
  \bibinfo{pages}{5115--5124}.
\newblock


\bibitem[\protect\citeauthoryear{Niepert, Ahmed, and Kutzkov}{Niepert
  et~al\mbox{.}}{2016}]%
        {niepert2016learning}
\bibfield{author}{\bibinfo{person}{Mathias Niepert}, \bibinfo{person}{Mohamed
  Ahmed}, {and} \bibinfo{person}{Konstantin Kutzkov}.}
  \bibinfo{year}{2016}\natexlab{}.
\newblock \showarticletitle{Learning convolutional neural networks for graphs}.
  In \bibinfo{booktitle}{\emph{International conference on machine learning}}.
  \bibinfo{pages}{2014--2023}.
\newblock


\bibitem[\protect\citeauthoryear{Pan, Zhou, Cao, Liu, Lukose, Scholz, and
  Yang}{Pan et~al\mbox{.}}{2008}]%
        {pan2008one}
\bibfield{author}{\bibinfo{person}{Rong Pan}, \bibinfo{person}{Yunhong Zhou},
  \bibinfo{person}{Bin Cao}, \bibinfo{person}{Nathan~N Liu},
  \bibinfo{person}{Rajan Lukose}, \bibinfo{person}{Martin Scholz}, {and}
  \bibinfo{person}{Qiang Yang}.} \bibinfo{year}{2008}\natexlab{}.
\newblock \showarticletitle{One-class collaborative filtering}. In
  \bibinfo{booktitle}{\emph{2008 Eighth IEEE International Conference on Data
  Mining}}. \bibinfo{pages}{502--511}.
\newblock


\bibitem[\protect\citeauthoryear{Qiao, Shen, Zhang, Wang, and Yuille}{Qiao
  et~al\mbox{.}}{2018}]%
        {qiao2018deep}
\bibfield{author}{\bibinfo{person}{Siyuan Qiao}, \bibinfo{person}{Wei Shen},
  \bibinfo{person}{Zhishuai Zhang}, \bibinfo{person}{Bo Wang}, {and}
  \bibinfo{person}{Alan Yuille}.} \bibinfo{year}{2018}\natexlab{}.
\newblock \showarticletitle{Deep co-training for semi-supervised image
  recognition}. In \bibinfo{booktitle}{\emph{Proceedings of the European
  Conference on Computer Vision (ECCV)}}. \bibinfo{pages}{135--152}.
\newblock


\bibitem[\protect\citeauthoryear{Rendle and Freudenthaler}{Rendle and
  Freudenthaler}{2014}]%
        {rendle2014improving}
\bibfield{author}{\bibinfo{person}{Steffen Rendle} {and}
  \bibinfo{person}{Christoph Freudenthaler}.} \bibinfo{year}{2014}\natexlab{}.
\newblock \showarticletitle{Improving pairwise learning for item recommendation
  from implicit feedback}. In \bibinfo{booktitle}{\emph{Proceedings of the 7th
  ACM international conference on Web search and data mining}}.
  \bibinfo{pages}{273--282}.
\newblock


\bibitem[\protect\citeauthoryear{Rendle, Freudenthaler, Gantner, and
  Schmidt{-}Thieme}{Rendle et~al\mbox{.}}{2009}]%
        {BPRMF}
\bibfield{author}{\bibinfo{person}{Steffen Rendle}, \bibinfo{person}{Christoph
  Freudenthaler}, \bibinfo{person}{Zeno Gantner}, {and} \bibinfo{person}{Lars
  Schmidt{-}Thieme}.} \bibinfo{year}{2009}\natexlab{}.
\newblock \showarticletitle{{BPR:} Bayesian Personalized Ranking from Implicit
  Feedback}. In \bibinfo{booktitle}{\emph{{UAI}}}. \bibinfo{pages}{452--461}.
\newblock


\bibitem[\protect\citeauthoryear{Sakai}{Sakai}{2018}]%
        {sakai2018laboratory}
\bibfield{author}{\bibinfo{person}{Tetsuya Sakai}.}
  \bibinfo{year}{2018}\natexlab{}.
\newblock \showarticletitle{Laboratory Experiments in Information Retrieval:
  Sample Sizes, Effect Sizes, and Statistical Power}.
  \bibinfo{publisher}{Springer}.
\newblock
\urldef\tempurl%
\url{https://link.springer.com/book/10.1007/978-981-13-1199-4}
\showURL{%
\tempurl}


\bibitem[\protect\citeauthoryear{Shi, Gong, Ding, MaXiaoyu~Tao, and Zheng}{Shi
  et~al\mbox{.}}{2018}]%
        {shi2018transductive}
\bibfield{author}{\bibinfo{person}{Weiwei Shi}, \bibinfo{person}{Yihong Gong},
  \bibinfo{person}{Chris Ding}, \bibinfo{person}{Zhiheng MaXiaoyu~Tao}, {and}
  \bibinfo{person}{Nanning Zheng}.} \bibinfo{year}{2018}\natexlab{}.
\newblock \showarticletitle{Transductive semi-supervised deep learning using
  min-max features}. In \bibinfo{booktitle}{\emph{Proceedings of the European
  Conference on Computer Vision (ECCV)}}. \bibinfo{pages}{299--315}.
\newblock


\bibitem[\protect\citeauthoryear{Sun, Lin, and Zhu}{Sun et~al\mbox{.}}{2020}]%
        {ke2020multi}
\bibfield{author}{\bibinfo{person}{Ke Sun}, \bibinfo{person}{Zhouchen Lin},
  {and} \bibinfo{person}{Zhanxing Zhu}.} \bibinfo{year}{2020}\natexlab{}.
\newblock \showarticletitle{Multi-Stage Self-Supervised Learning for Graph
  Convolutional Networks on Graphs with Few Labeled Nodes}.
\newblock \bibinfo{journal}{\emph{Proceedings of the AAAI Conference on
  Artificial Intelligence}}  \bibinfo{volume}{34} (\bibinfo{year}{2020}),
  \bibinfo{pages}{5892--5899}.
\newblock


\bibitem[\protect\citeauthoryear{Sun, Yang, Zhang, Bozzon, Huang, and Xu}{Sun
  et~al\mbox{.}}{2018}]%
        {sun2018recurrent}
\bibfield{author}{\bibinfo{person}{Zhu Sun}, \bibinfo{person}{Jie Yang},
  \bibinfo{person}{Jie Zhang}, \bibinfo{person}{Alessandro Bozzon},
  \bibinfo{person}{Long-Kai Huang}, {and} \bibinfo{person}{Chi Xu}.}
  \bibinfo{year}{2018}\natexlab{}.
\newblock \showarticletitle{Recurrent knowledge graph embedding for effective
  recommendation}. In \bibinfo{booktitle}{\emph{Proceedings of the 12th ACM
  Conference on Recommender Systems}}. \bibinfo{pages}{297--305}.
\newblock


\bibitem[\protect\citeauthoryear{Wang and Zhang}{Wang and Zhang}{2007}]%
        {wang2007label}
\bibfield{author}{\bibinfo{person}{Fei Wang} {and} \bibinfo{person}{Changshui
  Zhang}.} \bibinfo{year}{2007}\natexlab{}.
\newblock \showarticletitle{Label propagation through linear neighborhoods}.
\newblock \bibinfo{journal}{\emph{IEEE Transactions on Knowledge and Data
  Engineering}} \bibinfo{volume}{20}, \bibinfo{number}{1}
  (\bibinfo{year}{2007}), \bibinfo{pages}{55--67}.
\newblock


\bibitem[\protect\citeauthoryear{Wang, Zhang, Wang, Zhao, Li, Xie, and
  Guo}{Wang et~al\mbox{.}}{2018a}]%
        {RippleNet}
\bibfield{author}{\bibinfo{person}{Hongwei Wang}, \bibinfo{person}{Fuzheng
  Zhang}, \bibinfo{person}{Jialin Wang}, \bibinfo{person}{Miao Zhao},
  \bibinfo{person}{Wenjie Li}, \bibinfo{person}{Xing Xie}, {and}
  \bibinfo{person}{Minyi Guo}.} \bibinfo{year}{2018}\natexlab{a}.
\newblock \showarticletitle{Ripplenet: Propagating user preferences on the
  knowledge graph for recommender systems}. In
  \bibinfo{booktitle}{\emph{Proceedings of the 27th ACM International
  Conference on Information and Knowledge Management}}.
  \bibinfo{pages}{417--426}.
\newblock


\bibitem[\protect\citeauthoryear{Wang, Zhang, Xie, and Guo}{Wang
  et~al\mbox{.}}{2018b}]%
        {wang2018dkn}
\bibfield{author}{\bibinfo{person}{Hongwei Wang}, \bibinfo{person}{Fuzheng
  Zhang}, \bibinfo{person}{Xing Xie}, {and} \bibinfo{person}{Minyi Guo}.}
  \bibinfo{year}{2018}\natexlab{b}.
\newblock \showarticletitle{DKN: Deep knowledge-aware network for news
  recommendation}. In \bibinfo{booktitle}{\emph{Proceedings of the 2018 world
  wide web conference}}. \bibinfo{pages}{1835--1844}.
\newblock


\bibitem[\protect\citeauthoryear{Wang, Zhang, Zhang, Leskovec, Zhao, Li, and
  Wang}{Wang et~al\mbox{.}}{2019b}]%
        {wang2019knowledge2}
\bibfield{author}{\bibinfo{person}{Hongwei Wang}, \bibinfo{person}{Fuzheng
  Zhang}, \bibinfo{person}{Mengdi Zhang}, \bibinfo{person}{Jure Leskovec},
  \bibinfo{person}{Miao Zhao}, \bibinfo{person}{Wenjie Li}, {and}
  \bibinfo{person}{Zhongyuan Wang}.} \bibinfo{year}{2019}\natexlab{b}.
\newblock \showarticletitle{Knowledge-aware graph neural networks with label
  smoothness regularization for recommender systems}. In
  \bibinfo{booktitle}{\emph{Proceedings of the 25th ACM SIGKDD International
  Conference on Knowledge Discovery \& Data Mining}}.
  \bibinfo{pages}{968--977}.
\newblock


\bibitem[\protect\citeauthoryear{Wang, Zhang, Zhao, Li, Xie, and Guo}{Wang
  et~al\mbox{.}}{2019c}]%
        {wang2019multi}
\bibfield{author}{\bibinfo{person}{Hongwei Wang}, \bibinfo{person}{Fuzheng
  Zhang}, \bibinfo{person}{Miao Zhao}, \bibinfo{person}{Wenjie Li},
  \bibinfo{person}{Xing Xie}, {and} \bibinfo{person}{Minyi Guo}.}
  \bibinfo{year}{2019}\natexlab{c}.
\newblock \showarticletitle{Multi-task feature learning for knowledge graph
  enhanced recommendation}. In \bibinfo{booktitle}{\emph{The World Wide Web
  Conference}}. \bibinfo{pages}{2000--2010}.
\newblock


\bibitem[\protect\citeauthoryear{Wang, Zhao, Xie, Li, and Guo}{Wang
  et~al\mbox{.}}{2019d}]%
        {wang2019knowledge}
\bibfield{author}{\bibinfo{person}{Hongwei Wang}, \bibinfo{person}{Miao Zhao},
  \bibinfo{person}{Xing Xie}, \bibinfo{person}{Wenjie Li}, {and}
  \bibinfo{person}{Minyi Guo}.} \bibinfo{year}{2019}\natexlab{d}.
\newblock \showarticletitle{Knowledge graph convolutional networks for
  recommender systems}. In \bibinfo{booktitle}{\emph{The World Wide Web
  Conference}}. \bibinfo{pages}{3307--3313}.
\newblock


\bibitem[\protect\citeauthoryear{Wang, He, Cao, Liu, and Chua}{Wang
  et~al\mbox{.}}{2019a}]%
        {KGAT}
\bibfield{author}{\bibinfo{person}{Xiang Wang}, \bibinfo{person}{Xiangnan He},
  \bibinfo{person}{Yixin Cao}, \bibinfo{person}{Meng Liu}, {and}
  \bibinfo{person}{Tat{-}Seng Chua}.} \bibinfo{year}{2019}\natexlab{a}.
\newblock \showarticletitle{{KGAT:} Knowledge Graph Attention Network for
  Recommendation}. In \bibinfo{booktitle}{\emph{{Proceedings of the 25th ACM
  SIGKDD International Conference on Knowledge Discovery \& Data Mining}}}.
  \bibinfo{pages}{950--958}.
\newblock


\bibitem[\protect\citeauthoryear{Wang, Xu, He, Cao, Wang, and Chua}{Wang
  et~al\mbox{.}}{2020}]%
        {KGPolicy20}
\bibfield{author}{\bibinfo{person}{Xiang Wang}, \bibinfo{person}{Yaokun Xu},
  \bibinfo{person}{Xiangnan He}, \bibinfo{person}{Yixin Cao},
  \bibinfo{person}{Meng Wang}, {and} \bibinfo{person}{Tat-Seng Chua}.}
  \bibinfo{year}{2020}\natexlab{}.
\newblock \showarticletitle{Reinforced Negative Sampling over Knowledge Graph
  for Recommendation}. In \bibinfo{booktitle}{\emph{Proceedings of The Web
  Conference 2020}}. \bibinfo{pages}{99--109}.
\newblock


\bibitem[\protect\citeauthoryear{Wu, Li, So, Wright, and Chang}{Wu
  et~al\mbox{.}}{2012}]%
        {wu2012learning}
\bibfield{author}{\bibinfo{person}{Xiao-Ming Wu}, \bibinfo{person}{Zhenguo Li},
  \bibinfo{person}{Anthony~M So}, \bibinfo{person}{John Wright}, {and}
  \bibinfo{person}{Shih-Fu Chang}.} \bibinfo{year}{2012}\natexlab{}.
\newblock \showarticletitle{Learning with partially absorbing random walks}. In
  \bibinfo{booktitle}{\emph{Advances in Neural Information Processing
  Systems}}. \bibinfo{pages}{3077--3085}.
\newblock


\bibitem[\protect\citeauthoryear{Ying, He, Chen, Eksombatchai, Hamilton, and
  Leskovec}{Ying et~al\mbox{.}}{2018}]%
        {ying2018graph}
\bibfield{author}{\bibinfo{person}{Rex Ying}, \bibinfo{person}{Ruining He},
  \bibinfo{person}{Kaifeng Chen}, \bibinfo{person}{Pong Eksombatchai},
  \bibinfo{person}{William~L Hamilton}, {and} \bibinfo{person}{Jure Leskovec}.}
  \bibinfo{year}{2018}\natexlab{}.
\newblock \showarticletitle{Graph convolutional neural networks for web-scale
  recommender systems}. In \bibinfo{booktitle}{\emph{Proceedings of the 24th
  ACM SIGKDD International Conference on Knowledge Discovery \& Data Mining}}.
  \bibinfo{pages}{974--983}.
\newblock


\bibitem[\protect\citeauthoryear{Yu, Ren, Sun, Gu, Sturt, Khandelwal, Norick,
  and Han}{Yu et~al\mbox{.}}{2014}]%
        {yu2014personalized}
\bibfield{author}{\bibinfo{person}{Xiao Yu}, \bibinfo{person}{Xiang Ren},
  \bibinfo{person}{Yizhou Sun}, \bibinfo{person}{Quanquan Gu},
  \bibinfo{person}{Bradley Sturt}, \bibinfo{person}{Urvashi Khandelwal},
  \bibinfo{person}{Brandon Norick}, {and} \bibinfo{person}{Jiawei Han}.}
  \bibinfo{year}{2014}\natexlab{}.
\newblock \showarticletitle{Personalized entity recommendation: A heterogeneous
  information network approach}. In \bibinfo{booktitle}{\emph{Proceedings of
  the 7th ACM International Conference On Web Search and Data Mining}}.
  \bibinfo{pages}{283--292}.
\newblock


\bibitem[\protect\citeauthoryear{Zhang, Yuan, Lian, Xie, and Ma}{Zhang
  et~al\mbox{.}}{2016}]%
        {zhang2016collaborative}
\bibfield{author}{\bibinfo{person}{Fuzheng Zhang},
  \bibinfo{person}{Nicholas~Jing Yuan}, \bibinfo{person}{Defu Lian},
  \bibinfo{person}{Xing Xie}, {and} \bibinfo{person}{Wei-Ying Ma}.}
  \bibinfo{year}{2016}\natexlab{}.
\newblock \showarticletitle{Collaborative knowledge base embedding for
  recommender systems}. In \bibinfo{booktitle}{\emph{Proceedings of the 22nd
  ACM SIGKDD ACM SIGKDD International Conference on Knowledge Discovery \& Data
  Mining}}. \bibinfo{pages}{353--362}.
\newblock


\bibitem[\protect\citeauthoryear{Zhao, Yao, Li, Song, and Lee}{Zhao
  et~al\mbox{.}}{2017}]%
        {zhao2017meta}
\bibfield{author}{\bibinfo{person}{Huan Zhao}, \bibinfo{person}{Quanming Yao},
  \bibinfo{person}{Jianda Li}, \bibinfo{person}{Yangqiu Song}, {and}
  \bibinfo{person}{Dik~Lun Lee}.} \bibinfo{year}{2017}\natexlab{}.
\newblock \showarticletitle{Meta-graph based recommendation fusion over
  heterogeneous information networks}. In \bibinfo{booktitle}{\emph{Proceedings
  of the 23th ACM SIGKDD International Conference on Knowledge Discovery \&
  Data Mining}}. \bibinfo{pages}{635--644}.
\newblock


\end{thebibliography}
